\documentclass[12pt,journal,draftcls,oneside,onecolumn]{IEEEtran}

\usepackage{times}
\usepackage{cite}
\usepackage{color}
\usepackage{psfrag}
\usepackage{amssymb}
\usepackage{amsmath}
\usepackage{setspace}
\usepackage{mathrsfs}
\usepackage{amsfonts}
\usepackage{epsfig}
\usepackage{pifont}
\usepackage{array}
\usepackage[T1]{fontenc}
\usepackage{txfonts}
\usepackage{moreverb}
\usepackage[colorlinks,bookmarksopen,bookmarksnumbered,citecolor=red,urlcolor=red]{hyperref}
\usepackage{subfigure}
\usepackage{ifthen}
\usepackage{ifpdf}
\usepackage{booktabs}
\usepackage{changebar}
\usepackage{tikz}
\usepackage{algorithmic}
\usepackage{multirow}

\makeatletter
\def\ifundefined{\@ifundefined}

\begin{document}

\title{Energy Self-Sustainability in Full-Spectrum 6G}

\author{Jie~Hu,~\IEEEmembership{Member,~IEEE,}, Qing~Wang,~\IEEEmembership{Member,~IEEE}, Kun~Yang,~\IEEEmembership{Senior Member,~IEEE}
        
\thanks{Jie Hu and Kun Yang are with the School of Information and Communication Engineering, University of Electronic Science and Technology of China, Chengdu, 611731, China, email: hujie@uestc.edu.cn.}
\thanks{Kun Yang is also with the School of Computer Science and Electronic Engineering, University of Esssex, Colchester, CO4 3SQ, UK, email: kunyang@essex.ac.uk}
\thanks{Qing Wang is with the Department of Software Technology, Delft University of Technology, the Netherlands, email: qing.wang@ieee.org.}
\thanks{The authors would like to thank the financial support of National Natural Science Foundation of China (NSFC), Grant No. U1705263 and No. 61971102.}
}

\maketitle

\begin{abstract}

Full-spectrum ranging from sub 6 GHz to THz and visible light will be exploited in 6G in order to reach unprecedented key-performance-indicators (KPIs). However, extraordinary amount of energy will be consumed by network infrastructure, while functions of massively deployed Internet of Everything (IoE) devices are limited by embedded batteries. Therefore, energy self-sustainable 6G is proposed in this article. First of all, it may achieve network-wide energy efficiency by exploiting cell-free and airborne access networks as well as by implementing intelligent holographic environments. Secondly, by exploiting radio-frequency/visible light signals for providing on-demand wireless power transfer (WPT) and for enabling passive backscatter communication, ``zero-energy'' devices may become a reality. Furthermore, IoE devices actively adapt their transceivers for better performance to a dynamic environment. This article aims to provide a first glance at primary designing principles of energy self-sustainable 6G.

\end{abstract}

\section{Introduction}

With 5G mobile systems being commercially rolling out gradually, research discussions into next-generation mobile systems, i.e., 6G, have started \cite{Latva-Aho2019}. 5G will incur 10 times energy consumption of 4G due to increased density of cells and that of antennas, though energy efficiency per bit has increased. With 6G going towards higher spectrum (such as Tera-Hertz or THz) and thus resulting in even denser networks and smaller cells, energy consumption will become a big hurdle on the way to 6G success. On the infrastructure side, a huge amount of energy will be consumed for powering numerous RF chains connected to a vast number of antennas, for extraordinary wideband signal processing, for maintaining a satisfactory coverage and for tracking mobile devices with super-narrow beams. Therefore, reducing energy consumption and jointly coordinating distributed infrastructure to achieve network-wide optimal energy efficiency constitute the first challenge in future 6G.

On the other hand, the Internet of Everything (IoE), as envisioned as a major 6G application, means that a vast number of small devices will be connected to networks. These devices are typically either battery-powered or battery-less. However, conventional RF chain enabled communication as well as handshaking based access are both battery killers. In order to extend life-cycle of IoE devices, their energy harvesting capabilities from the ambient environment have to be activated. However, the communication performance of IoE devices is largely constrained by intermittent energy supplies. Therefore, designing super-low-power IoE transceivers and enabling an on-demand and cost-effective manner of energy supply to these IoE devices constitute the second challenge in future 6G.

We envision 6G mobile systems have to be energy self-sustainable, both at the infrastructure side (such as traditional base stations) and at the device side (may it be in the form of smart phones or implanted brain computer interface devices), in order to be deployed at large and to achieve commercial success. There is a consensus that visible light (VL) communications will be an integral part of 6G in addition to the traditional medium/spectrum of radio frequency (RF) due to VL's unique properties like massive bandwidth, information security, economical, no harm to humans, etc. This paper serves to bring these two currently separate research domains, i.e., RF and VL, together to provide a full-spectrum 6G mobile system. A joint design that inherently considers both RF and visible light, shall be the methodology to be taken from the onset and in particular with energy self-sustainability borne in mind.

Rather than reinventing new technologies into the already crowded family of 6G enabling technologies, this paper aims to explore how the existing members of this family such as THz, VL, intelligence reflecting surface (IRS) and AI can be explored for the energy self-sustainability. And this exploration needs to be conducted in a systematic and holistic manner at the design stage of 6G. Fortunately, each of these 6G candidate technologies provides promising potentials to empower energy self-sustainability. This paper endeavours to shed some light on potential solutions to energy supply issues in 6G with the hope to spark more discussion and research into this critical aspect of 6G mobile systems. It is this article's belief that 6G systems need to be energy self-sustainable and this feature decides 6G's wide commercial success in the future. 

For this purpose, this article firstly organizes the key 6G enabling technologies into a hierarchical system architecture for future 6G mobile systems, as presented in Section II. Without losing generosity, the presentation of the architecture is more tilted towards the angle of energy supply, which is the focus of this article. Guided by this architecture, mechanisms to potentially enable energy self-sustainability are discussed from two aspects: infrastructure side and IoE device side, respectively in Section III and IV. The infrastructure side focuses on the distributed access network layer with particular focus on three new features of 6G, namely, cell-free access networks and airborne access networks (the organization of the network) as well as smart surface (the composing materials of the network). On the IoE device side, discussions are given to high-resolution signal processing based wireless information and power transfer (WIPT), multifunctional IoE devices and human-in-the-loop based transceiver adaptation. 

\begin{figure}[t]
	\centering
  	\includegraphics[width=0.9\linewidth]{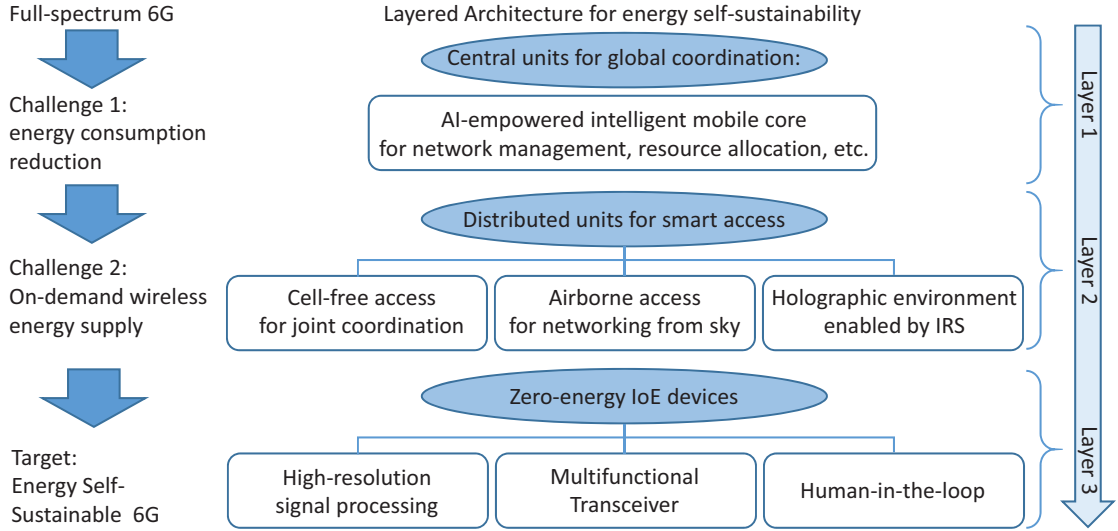}
  	\caption{Technical challenges and solutions towards energy self-sustainable 6G.}\label{fig:Structure}
  	\setlength{\belowcaptionskip}{0pt}
  	\setlength{\abovecaptionskip}{0pt}
\end{figure}

The contribution of the article lies in twofold. Firstly, it proposes a 6G mobile system architecture for the first time where the traditional base stations are decomposed into two complementing layers: central units (CUs) and distributed units (DUs). CUs are analogous to today's base stations in terms of their deployment locations but with more intelligence. DUs reside at very close proximity to end devices due to the employment of much higher frequency spectrum in 6G. DUs may be formed as a small cell-free network and using IRS. Secondly, these 6G enabling technologies are reinvestigated together with other new proposals, from the angle of energy self-sustainability by giving insightful discussions towards a commercially viable future 6G mobile systems. These investigations are centred around two complementing aspects: reducing energy consumption (i.e., energy efficiency) and opening new energy supplies. The technical challenges and solutions towards energy self-sustainable 6G are summarised in Fig. \ref{fig:Structure}.

\section{Architecture of Energy self-sustainable 6G}

\begin{figure}[t]
	\centering
  	\includegraphics[width=0.9\linewidth]{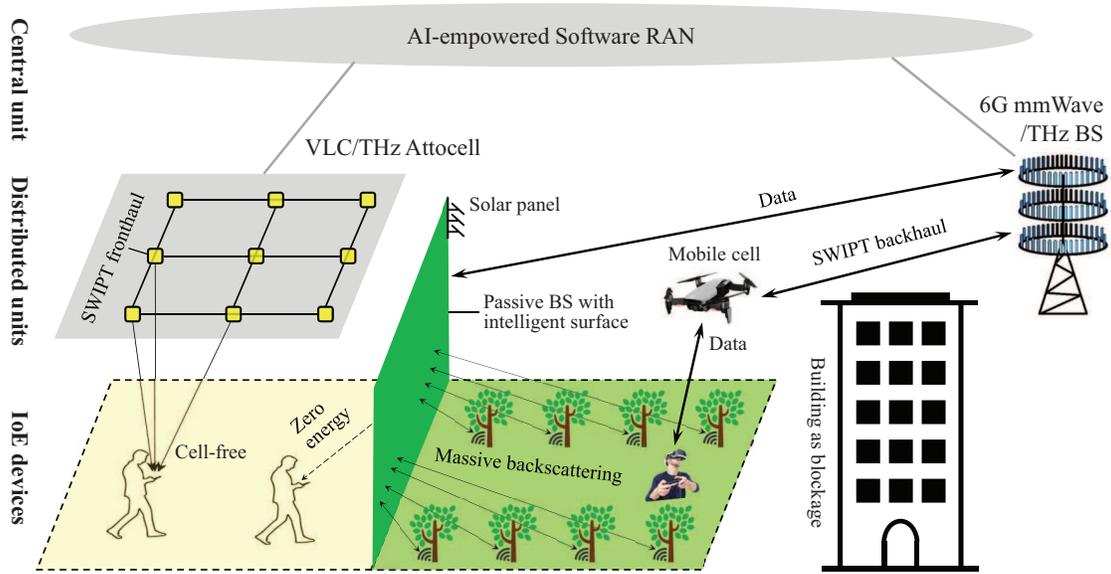}
  	\caption{Architecture of energy self-sustainable 6G network}\label{fig:Arch}
  	\setlength{\belowcaptionskip}{0pt}
  	\setlength{\abovecaptionskip}{0pt}
\end{figure}

In order to achieve more stringent quality of service (QoS), 6G has to move up to even higher spectral bands than mmWave, such as THz and visible light band. However, 6G may still utilise sub-6 GHz and mmWave band, which are the legacy of 5G. Therefore, 6G is foreseen as a \emph{full-spectrum} mobile communication system. Different spectrum will be exploited for various objectives and applications: $1)$ \textit{Sub-6 GHz band} is used for large coverage and seamless handover; $2)$ \textit{mmWave band} is leveraged both for wireless backhaul and fronthaul to fixed targets and for satisfying high-rate requirement of mobile targets; $3)$ \textit{THz band} is used to provide ultra-high data rate and low-latency service for applications such as pervasive XR; $4)$ \textit{VL band} with extra bandwidth can be further exploited for delivering holographic tele-presence with other support of high-resolution imaging and sensing. VL cannot penetrate blockage and it is capable of creating a secure wireless communication.
	
In this work, we focus on the access part of full-spectrum 6G networks. Our proposed energy self-sustainable architecture for 6G networks, as shown in Fig.~\ref{fig:Arch}, consists of three layers: 1) Central unit (CU), e.g., AI-empowered software RAN, 2) Distributed units (DUs), e.g. self-organised attocells and 3) zero-energy IoE device.
The AI-empowered CU determines which spectrum and which DUs are used to serve a group of IoE devices. 
Based on the dynamics at the IoE devices and surrounding environments, the DUs form the exact cells to serve IoE devices.
Each layer in the architecture, from different dimensions, can contribute to the energy sustainability of future 6G networks.

The target of a CU is to provide demanded services to the IoE devices in an energy-efficient manner and/or with green energy (e.g., produced by nearby solar panel), all contributing to the network energy self-sustainability. For example, if the IoE device is within the indoor area where dense visible light or THz attocells are deployed, the IoE device can be served by these attocells which have the capability to provide both fast, ultra-reliable, and energy-efficient wireless connectivity. If the IoE device is approaching a passive DU that is equipped with intelligent surfaces, the IoE device could be guided by the AI-empowered central unit to switch to the cell of the passive DU, which could provide low- or even zero-energy wireless connectivity. If an outdoor IoE device requires high-speed wireless connections for applications such as XR, while in the surroundings a fixed-location and high-rate DU is not available, the AI-power CU can schedule a high-rate mobile DU to serve the IoE device. Drones or mobile robots can host the mobile DU. The backhaul enables WIPT to mobile DUs for achieving energy self-sustainability. Due to the complexity of decision making, new AI techniques such as deep reinforcement learning (DRL) is a necessity for problem solving. To avoid overwhelming the CU, some processing and decision making will be conducted at DUs under the coordination of the CU. A right balance between the CU and DU processing is a multi-objective optimization problem that may need DRL. Software defined networking (SDN) and network function virtualization (NFV) will continue to be evolved and then employed to dynamically deploy new QoS (Quality of Service) algorithms or controller on the fly.
	
For most of the 6G DUs, we envision that a massive number of distributed antennas will be densely deployed to achieve both ultra-high reliability and wide coverage. As a result, many ground-breaking techniques could be enabled and leveraged in 6G. One of them is cell-free networking that can provide ultra-reliable and energy-efficient connectivity \cite{DenseVLC2020}. Also, cells formed by the DUs must adapt to surrounding dynamics to perform energy-efficient communication. For example, the attocells must react to dynamic blockages in the surroundings.
	
IoE devices play essential roles in an energy self-sustainable 6G network. First, zero-energy IoE devices, which are powered by wideband WIPT, could be primarily deployed to provide massive connections and to realize IoE. For example, a large number of backscattering nodes can be connected to IRS enabled DUs to create an Internet of Trees, as illustrated in Fig.~\ref{fig:Arch}. Second, in our proposed architecture for future energy self-sustainable 6G networks, it is essential for IoE devices to adapt to the surrounding environment. The adaptation can be performed without the involvement of human beings, such as physically adjusting the receiving antennas to a better position to achieve higher data rates with less energy. Another possibility is the involvement of human beings. With wireless communications moving to THz and VL, the human body can easily block the surrounding wireless links. Therefore, the future 6G network could quickly become more energy-efficient if it guides human beings to position their bodies to recoup the blocked wireless communication links.

\section{Distributed Units for Smart Access} 

Towards energy self-sustainability, every Joule of energy in 6G should be efficiently consumed in access networks by adopting the following methods: 1) Breaking cellular boundaries for globally scheduling DUs to deliver seamless services and to achieve network-wide energy efficiency; 2) Deploying flexible airborne access network to increase energy efficiency by further considering the vertical dimension and to realise wireless charging for battery-powered or batteryless devices from the sky; 3) Actively changing adverse wireless environment to be beneficial by implementing IRSs so as to substantially increase the efficiency of WIPT. 

\subsection{Cell-Free Access}

In a cell-free massive MIMO (CFM-MIMO) system, a large number of DUs are connected to a CU via wired/wireless fronthaul. Joint coordination among these DUs results in cell-free access networks, where IoE devices are simultaneously served by cooperative BSs. Therefore, handover in a cell-free access network is different, since cellular boundaries are broken. Furthermore, joint coordination of a CU may achieve a network-wide energy efficiency \cite{Jiang2018}. However, full-cooperation among DUs may result in heavy tele-traffic load on fronthaul, since it requires the exchange of network state information, the transmission of transmit beamformers and requested information.

In higher spectral bands of 6G, such as THz and VL, signal propagation suffers from significant penetration loss, if a lot of opaque obstacles are distributed in wireless environments. By further considering the increasing path-loss in high frequency bands, super-dense DUs have to be deployed for reducing distances and for ensuring line of sight (LoS) between DUs and IoE devices. However, more energy has to be consumed for powering such large-scale access network, while increased interference imposed by super-dense DUs has to be carefully managed and controlled. Furthermore, low-cost DUs can be equipped with solar panels for harvesting energy from ambient environment. They can be connected to smart grid to enable energy trade among their peers. As a result, global energy cooperation may result in an improved network-wide energy efficiency.

As the number of DUs grows exponentially in cell-free access networks of 6G, wireless fronthaul is indispensable since the expense of deploying wired fronthaul cannot be afforded. RF signal based WIPT \cite{8421584} is an economic approach to transfer both information and energy from the CU to low-cost DUs via fronthaul in order to maintain energy self-sustainability. Due to hardware limitations, we only have limited RF chains at a THz based transmitter, which is far lower than the number of antennas compacted. Therefore, the accuracy of transmit beams cannot be guaranteed. Energy efficient hybrid transceiver consisting of both digital and analog beamformers is compromised for trying to achieve closest performance to the ideal full-digital solution \cite{8733134}. Moreover, in both THz and VL based cell-free access network, the assignment of DUs and that of antennas to IoE devices can be formulated as a network-wide energy efficiency maximisation problem, while satisfying various QoS requirements of devices, suppressing interference imposed by large-scale DUs and their associated antennas.

Handover of mobile devices in cell-free access networks with densely deployed DUs faces novel challenges. Since the number of mobile devices is much lower than that of DUs, user-centric networking is the primary principle in a cell-free access network. Different from conventional handover, where mobile devices traverse across cells, a virtual cell dynamically formed by cooperative DUs tracks the movement of a mobile device, which constitutes a mobile cell serving a specific mobile device. Therefore, the formation of a dynamic virtual cell needs to be completed in real-time in order to provide seamless service experience, which calls for low-complexity heuristic algorithms with near-optimal performance. By exploiting up-to-date location information of mobile devices, we may significantly reduce the complexity of algorithms and increase the energy efficiency for virtual cell formation as well as for DUs/antennas assignment. 

Moreover, in a cell-free access network, synchronising densely deployed DUs is also essential. By exploiting the deterministic reflection of THz waves and VL, we may achieve over-the-air synchronisation among all the DUs.

\subsection{Airborne Access} 

\begin{figure}[!t]
	\centering
	\includegraphics[width=0.8\linewidth]{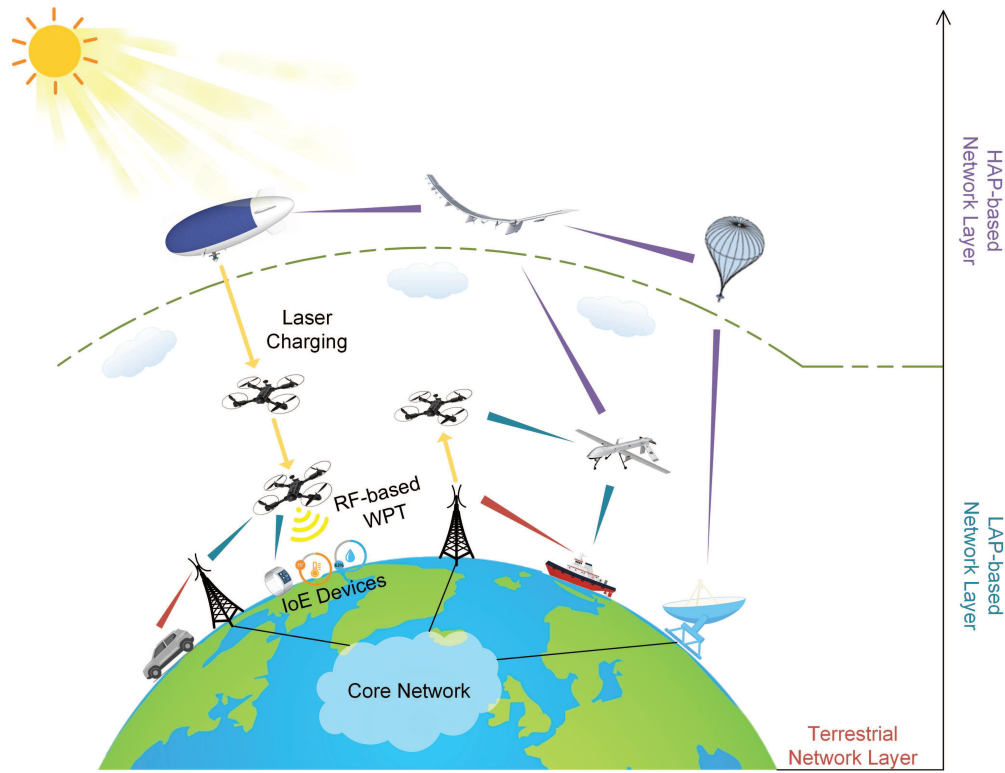}
	\caption{Hierarchical aerial access network}
	\label{fig:ARAN}
\end{figure}

Different sorts of aircraft constitute hierarchical airborne access networks (AANs) by carrying RF/VL transceivers on board, as illustrated in Fig.~\ref{fig:ARAN}. Generally, based on their flying height, aircraft in AANs are classified into high-altitude platforms (HAPs) (e.g., high-altitude balloons and aircrafts) and low-altitude platforms (LAPs) (e.g., drones). AANs have to be energy self-sustainable, since we cannot provide stable power supply from the ground. Joint coordination among heterogeneous aircraft is capable of extending coverage of air-to-ground communication and of realising flexible network deployment as well as of providing efficient WIPT services.

LAPs are mainly constituted by energy-constraint devices powered by embedded batteries, as portrayed in Fig.~\ref{fig:ARAN}. Therefore, 3D trajectories and flight control should be carefully designed in order to reduce energy consumption of aircraft, while achieving efficient-efficient downlink wireless power transfer (WPT) services, downlink/uplink wireless information transfer (WIT) services and computation offloading services to battery-powered or batteryless ground devices \cite{8805125}. Moreover, the energy consumption of ground devices can be substantially reduced, since flying LAPs reduce signal propagation distances to them. In addition, laser charging is an efficient way to power LAPs so as to extend their flight time, since it may focus a high amount of energy in a narrow beam. Laser beams are normally formed by ground stations or aircraft in HAPs.

HAPs can be aloft in the air for a relatively long time, as illustrated in Fig.~\ref{fig:ARAN}. A key advantage of HAPs is their self-adjustable positions in order to maintain efficient WIPT towards LAPs. Moreover, HAPs can harvest stable solar energy for powering their engines and transceivers, since they are floating in the stratosphere \cite{9013552}.

Careful coordination among aircraft in different layers may result in a network-wide optimal energy efficiency and it may provide significant performance gain in the vertical dimension of energy self-sustainable 6G.

\subsection{Holographic Environment}

Radiated RF/VL signals are weakened as they propagate in wireless channels, which may substantially degrade both WIPT performance. They may be absorbed, reflected and scattered by obstacles distributed in wireless environment. Signals from different propagation paths may be destructively combined, which results in substantial fading. Wireless channels become even worse in THz and VL band, since signals can be easily absorbed by large molecules and they can only be reflected \cite{Sheikh2020}. Therefore, in order to overcome more serious channel attenuation in THz and VL band, line-of-sight transmission path has to be guaranteed between a transmitter and receiver pair, while more energy has to be consumed for achieving satisfactory WIPT performance.  

\begin{figure}[t]
	\centering
  	\includegraphics[width=0.8\linewidth]{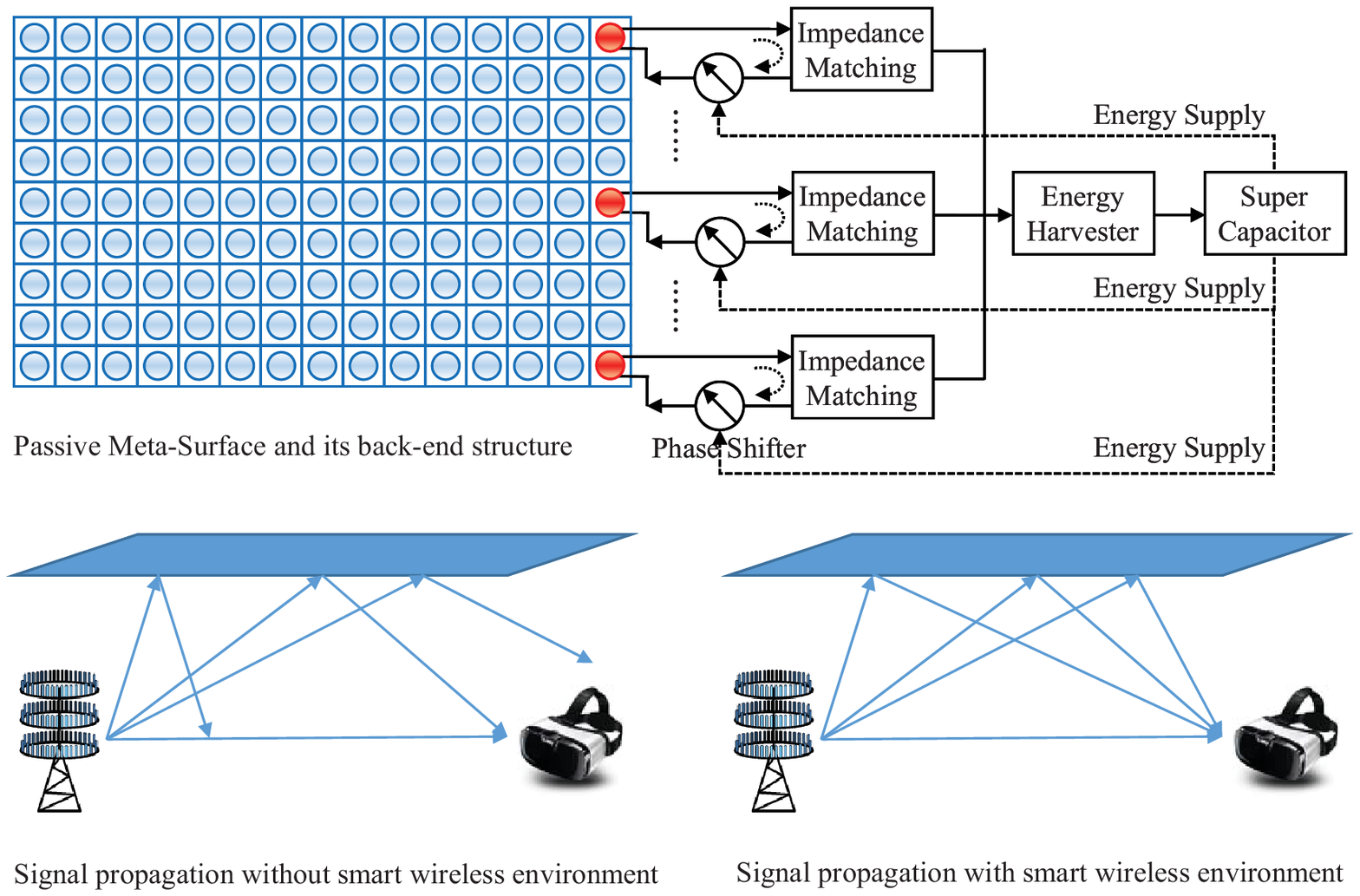}
  	\caption{Basic back-end structure of IRSs and signal propagation with/without smart wireless environment.}\label{fig:Surfaces}
  	\setlength{\belowcaptionskip}{0pt}
  	\setlength{\abovecaptionskip}{0pt}
\end{figure}

In all classic communication systems, only transmission or receiving strategies at transceivers can be designed for efficient WIPT, such as beamforming, adaptive modulation and coding design at transmitters as well as signal combining and iterative decoding design at receivers, which all aim for counteracting signal attenuation and fading in wireless channels. This is because characteristics of wireless channels cannot be changed as exemplified in Fig. \ref{fig:Surfaces}.

As electronic materials rapidly progress, IRSs can be exploited for actively changing characteristics of wireless channels. For example, based on signals' frequencies, phases and amplitudes, IRss can be intelligently adjusted, so that signals reflected by them can be constructively combined at receivers, as illustrated in Fig. \ref{fig:Surfaces}. As a result, we may artificially create a line-of-sight (LoS) transmission path in THz and VL band, when no actual LoS path exists between a transmitter and receiver pair. Therefore, received signal strength can be substantially increased without extra energy consumption, which results in more energy efficient WIPT.

IRSs are not equipped with any active RF chains, while they are not connected to any stable energy sources. As shown in Fig. \ref{fig:Surfaces}, a 2-dimensional IRS consists of a range of passive reflectors. Signals received by a reflector are then transferred to a programmable impedance matching network (IMN). By adjusting electronic components in this IMN, we may arbitrarily adjust its reflection coefficient. Therefore, A portion of received signals penetrate and their energy is harvested and stored in super-capacitors, as illustrated in Fig. \ref{fig:Surfaces}. The other portion are reflected via a programmable analog phase-shifter. Carefully designed phase-shifters may result in a constructive combination of reflected signals at the receiver. Note that programmable analog phase shifters and IMN have to be powered by energy harvested. Therefore, it is essential to design reflection coefficients for all reflectors. Higher reflection coefficients generate more strong reflected signals but limited energy can be harvested by the surface. Given insufficient energy harvested, we are only able to adjust several phase-shifters not all. Therefore, we have to jointly select which phase-shifters to be activated and how they change the phases of reflected signals in order to maximise the signal strength at the receiver. By contrast, lower reflection coefficients result in more energy harvested by the surface. Therefore, more phase-shifters can be activated, which results in well-tuned reflected signals. However, the strengths of the reflected signals are weak, which may degrade WIPT performance.

In a nutshell, given an IRS, we need to jointly design reflection coefficients of all the reflectors, activation of phase-shifters and how they tune reflected signals. Furthermore, transmit beamforming at transmitters, intelligent reflecting at IRSs and receive combining at receivers can be jointly designed to create a holographic wireless environment in energy self-sustainable 6G \cite{Holographic}.

\section{Zero-Energy IoE Devices}

In energy self-sustainable 6G, battery-powered or batteryless IoE devices requires controllable and on-demand WPT for maintaining seamless service experience, while their energy consumption has to be kept as low as possible. Moreover, these IoE devices should also intelligently adapt their transceivers to environments for achieving more efficient information reception and energy harvesting.  

\begin{figure}[t]
	\centering
  	\includegraphics[width=0.9\linewidth]{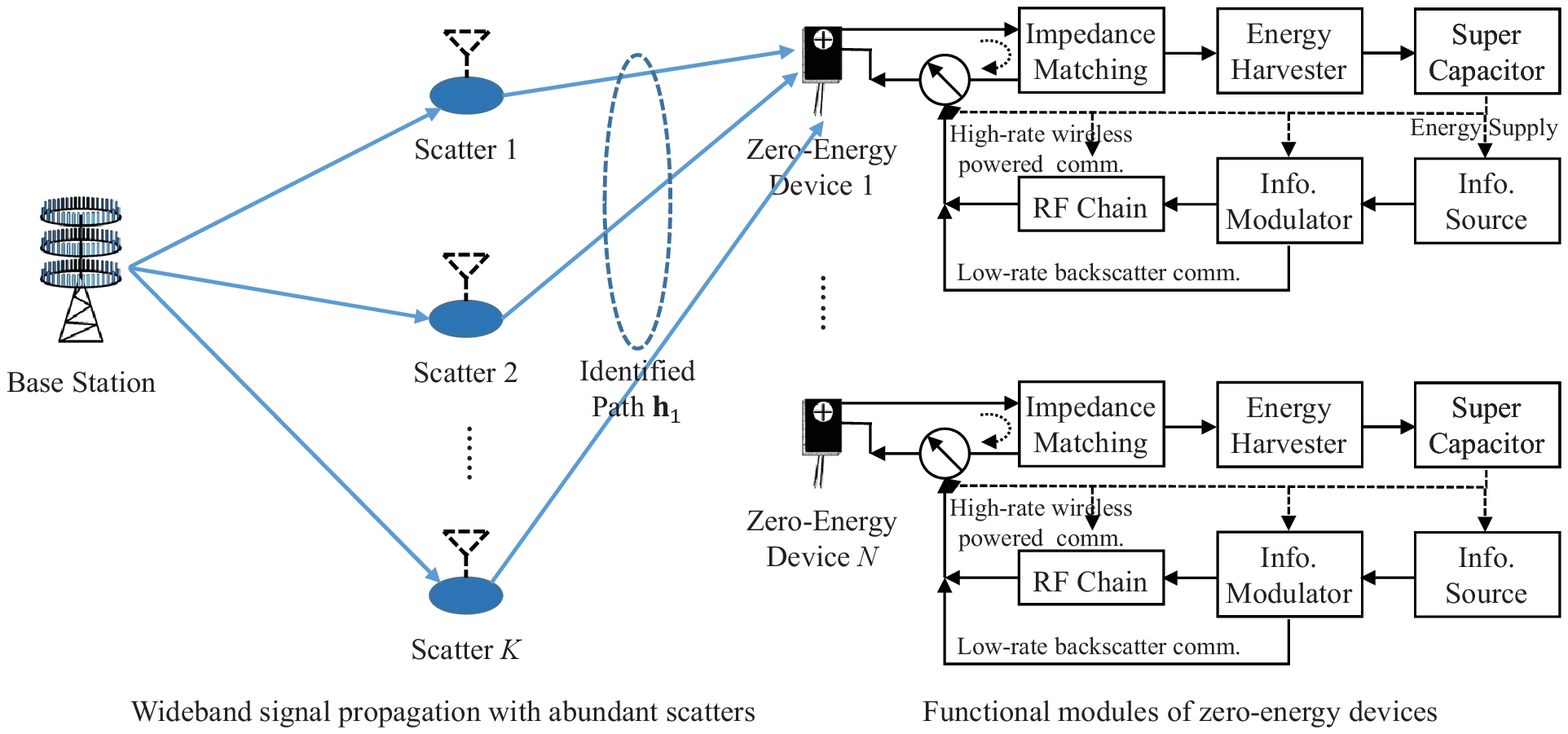}
  	\caption{Wideband signal propagation and functional modules of zero-energy devices}\label{fig:Zero-Energy}
  	\setlength{\belowcaptionskip}{0pt}
  	\setlength{\abovecaptionskip}{0pt}
\end{figure}

\subsection{High-resolution Signal Processing based WIPT}

RF signals can be exploited for flexible, controllable and on-demand WPT in any time and anywhere. On infrastructure side, we do not need any additional hardware implementation. Unmodulated and modulated RF signals can be relied upon for WIPT. We have to fully exploit characteristics of wideband signals for WIPT in energy self-sustainable 6G.

There are full of scatters and reflectors in the propagation of RF/VL signals, which results in multi-path transmissions from a transmitter to a receiver. The number of identified paths at the receiver is determined by the following factors:
\begin{itemize}
	\item Signal power $P_t$: a higher signal power results in a higher signal-to-noise-ratio (SNR) and clearer channel impulsive responses at the receiver.
	\item Signal bandwidth $W$: If the signal propagation delay between two transmission paths is higher than the coherence time $1/W$, this pair of paths cannot be identified. Otherwise, the receiver is capable of identifying these two transmission paths. 
	\item Geographic positions of scatters and reflectors: Given a pair of transmission paths, if their distance have a difference of $c/W$, where $c$ is the speed of light, they can be identified at the receiver, otherwise they cannot. 
\end{itemize}
Therefore, if we have an extremely high bandwidth $W$, more scatters and reflectors may generate more identified transmission paths \cite{Xu2018}. We do not rely upon coarse-grained signal processing on the basis of antenna-to-antenna. Instead, wide-bandwidth transmission in 6G creates a chance of high-resolution signal processing on the basis of identified transmission paths, which provides substantial spatial multiplexing and diversity gains for WIPT, as illustrated in Fig. \ref{fig:Zero-Energy}. By fully exploiting this additional degree of freedom in signal design and the resultant performance gain, we may also substantially reduce energy consumption of transmitters. 

\subsection{Multifunctional Transceiver}

IoE devices should have ability of gleaning energy from downlink RF/VL signals, while their energy consumption has to be suppressed. Therefore, as portrayed in Fig. \ref{fig:Zero-Energy}, the following functional modules have to be implemented at IoE devices: 1) an IMN, which is a circuit to guarantee the alternative current (AC) energy carried by the received RF signals can be delivered to the back-end by matching the impedance of antennas and that of back-end's circuits; 2) an energy harvester, which is a circuit for rectifying AC energy carried by RF/VL signals to direct current (DC) in order to drive electronic loads or to charge energy storage units. Key electronic elements in rectifiers are diodes; 3) an energy storage unit can be either a battery or a super capacitor. Batteries have long-term energy storage capabilities, but its charging efficiency is low. Super-capacitors can only store energy for a short-term but its charging efficiency is very high.

Corresponding to different QoS, we have the following pair of designs for batteryless IoE devices:
\begin{itemize}
	\item High-rate wireless powered communication: IoE devices adopt a harvest-store-then-transmit protocol. In order to increase energy harvesting efficiency, we should guarantee a perfect impedance matching. Only when they have sufficient energy in their storage units, their active information transmissions commence via RF/VL chains. Therefore, IoE devices can obtain high-rate information transmissions \cite{7982605}. However, powering RF/VL chains is an energy-consuming task.
	\item Low-rate backscatter communication: IoE devices deliberately mismatch the impedance of the receiving antennas and the back-end circuits. Therefore, RF/VL signals received by antennas cannot penetrate to the back-end but backscattered instead. IoE devices can flexibly modulate their own information on backscattered signals in frequency and time domains. Therefore, IoE devices do not need any RF/VL chains for information transmission, which may substantially reduce energy consumption \cite{10.1145/3300061.3345451}. Furthermore, we can fully exploit the abundant full-spectrum of 6G by intelligently changing the frequencies of backscattered signals in order to avoid congested bands. However, IoE devices also need to harvest energy for powering their signal processing modules. Therefore, backscattering coefficients need to be carefully designed for allowing some of the received RF/VL signals flowing into the back-end for energy harvesting. The IMN of IoE devices should also be programmable.
\end{itemize}
Furthermore, we may integrate the above-mentioned designs into a single IoE, as exemplified in Fig. \ref{fig:Zero-Energy}. Therefore, low-rate control signalling or sporadic information transmissions can be completed by passive backscatter communication, while high-rate data transmissions can be completed by active wireless powered communication. 

\subsection{Human-in-the-loop}

Signal propagation in these bands heavily relies on LoS between transmitters and receivers. If no LoS link exists, WIPT performance would severely degrade. In both indoor and outdoor scenarios, many objects in surrounding wireless environments may block LoS links, such as walls, cabinets, moving objects and etc. Moreover, human bodies are mostly neglected as critical blockages in the design of a communication system operating in high frequency bands, especially when hand-held communication devices are taken into account. 

\begin{figure}[t]
	\centering
	\includegraphics[width=0.9\columnwidth]{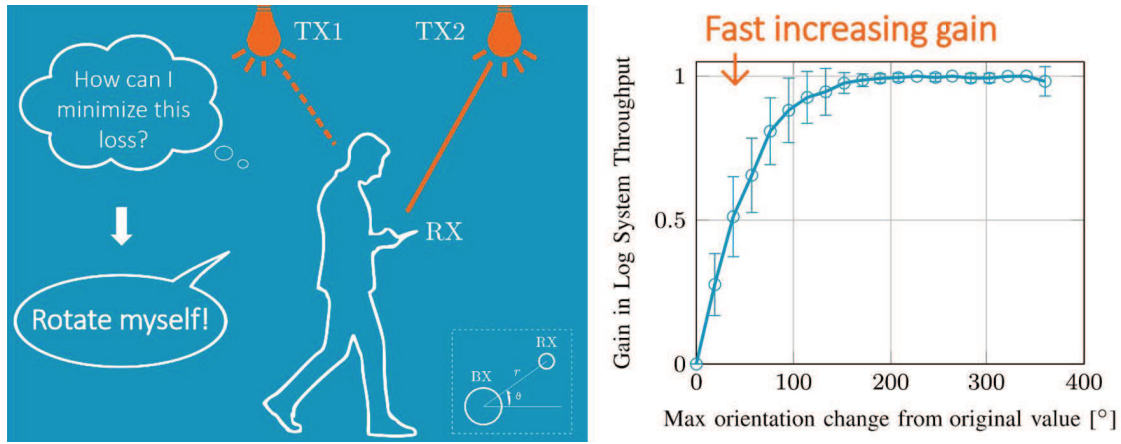}
	\caption{Human-in-the-loop adaptation to the surrounding environment (BX: blockage)}
	\label{fig:adaptationtoenvironment}
\end{figure}

In order to improve WIPT performance, it is crucial for IoE devices to possess abilities of intelligently adapting their transceivers to dynamic wireless environments. This adaptation can be achieved without/with intervention from human users. For instance, reconfigurable antennas of IoE devices can be automatically adjusted to better positions in order to gain a higher information rate and to harvest more energy. Interactions between devices and human users may also result in environmental adaptation. These interactions are known as ``human-in-the-loop''. As illustrated in Fig.~\ref{fig:adaptationtoenvironment}, human users may rotate their bodies to enable LoS links. In a case study of VL based communication network, observe from the right part of Fig.~\ref{fig:adaptationtoenvironment} that allowing a rotation angle of $76^{\circ}$ achieve a system throughput gain of $81\%$~\cite{NutVLC}. The involvement of human users in transceiver adaptation becomes normal in our daily life. Human users holding devices always try to rotate their bodies to gain high information rates. This is more efficient in VL band of future 6G, since VL can be easily captured by human eyes. Human users may readily aim their devices to intensive VL.  

Autonomous transceiver adaptations naturally find wireless channels having lowest channel attenuation. Therefore, energy consumed by infrastructure to radiate RF/VL signals can be significantly reduced. Based on surrounding wireless environments, IoE devices may decide how to adapt themselves to reduce energy consumption in information transmission and reception as well as to increase energy harvested in downlink WPT, which is also a key step towards zero-energy devices.

\section{Conclusion}

This article provides a layered architecture of energy self-sustainable 6G in full-spectrum which consists of AI-empowered CUs, massively deployed DUs and IoE devices. In order to deal with a tremendous amount of energy consumed for satisfying unprecedented QoS requirements, key solutions are provided to cell-free access for joint coordination of DUs, airborne access for 3D networking and holographic environmental design enabled by IRSs for counteracting severe channel attenuation in THz and VL. Furthermore, we provide vital approaches to realise zero-energy IoE devices, such as high-resolution signal processing based WIPT, multifunctional transceivers without any batteries and active transceiver adaptation to dynamic environments. However, there are still numerous challenges of making energy self-sustainable 6G a practice, which calls for a joint effort from both academia and industry.

\bibliography{Reference}

\begin{thebibliography}{10}
\providecommand{\url}[1]{#1}
\csname url@samestyle\endcsname
\providecommand{\newblock}{\relax}
\providecommand{\bibinfo}[2]{#2}
\providecommand{\BIBentrySTDinterwordspacing}{\spaceskip=0pt\relax}
\providecommand{\BIBentryALTinterwordstretchfactor}{4}
\providecommand{\BIBentryALTinterwordspacing}{\spaceskip=\fontdimen2\font plus
\BIBentryALTinterwordstretchfactor\fontdimen3\font minus
  \fontdimen4\font\relax}
\providecommand{\BIBforeignlanguage}[2]{{%
\expandafter\ifx\csname l@#1\endcsname\relax
\typeout{** WARNING: IEEEtran.bst: No hyphenation pattern has been}%
\typeout{** loaded for the language `#1'. Using the pattern for}%
\typeout{** the default language instead.}%
\else
\language=\csname l@#1\endcsname
\fi
#2}}
\providecommand{\BIBdecl}{\relax}
\BIBdecl

\bibitem{Latva-Aho2019}
M.~Latva-Aho and K.~Lepp{\"{a}}nen, \emph{{Key Drivers and Research Challenges
  for 6G Ubiquitous Wireless Intelligence - 6G Research Visions 1, September
  2019}}, 2019, no. September.

\bibitem{DenseVLC2020}
J.~{Beysens}, Q.~{Wang}, A.~{Galisteo}, D.~{Giustiniano}, and S.~{Pollin}, ``{A
  Cell-Free Networking System With Visible Light},'' \emph{{IEEE/ACM
  Transactions on Networking}}, pp. 1--16, 2020.

\bibitem{Jiang2018}
R.~Jiang, Q.~Wang, H.~Haas, and Z.~Wang, ``Joint user association and power
  allocation for cell-free visible light communication networks,'' \emph{IEEE
  Journal on Selected Areas in Communications}, 2018.

\bibitem{8421584}
J.~{Hu}, K.~{Yang}, G.~{Wen}, and L.~{Hanzo}, ``Integrated data and energy
  communication network: A comprehensive survey,'' \emph{IEEE Communications
  Surveys Tutorials}, vol.~20, no.~4, pp. 3169--3219, Fourthquarter 2018.

\bibitem{8733134}
S.~A. {Busari}, K.~M.~S. {Huq}, S.~{Mumtaz}, J.~{Rodriguez}, Y.~{Fang}, D.~C.
  {Sicker}, S.~{Al-Rubaye}, and A.~{Tsourdos}, ``Generalized hybrid beamforming
  for vehicular connectivity using thz massive mimo,'' \emph{IEEE Transactions
  on Vehicular Technology}, vol.~68, no.~9, pp. 8372--8383, Sep. 2019.

\bibitem{8805125}
Y.~{Du}, K.~{Yang}, K.~{Wang}, G.~{Zhang}, Y.~{Zhao}, and D.~{Chen}, ``Joint
  resources and workflow scheduling in uav-enabled wirelessly-powered mec for
  iot systems,'' \emph{IEEE Transactions on Vehicular Technology}, vol.~68,
  no.~10, pp. 10\,187--10\,200, Oct 2019.

\bibitem{9013552}
Y.~{Du}, K.~{Wang}, K.~{Yang}, and G.~{Zhang}, ``Trajectory design of
  laser-powered multi-drone enabled data collection system for smart cities,''
  in \emph{2019 IEEE Global Communications Conference (GLOBECOM)}, Dec 2019,
  pp. 1--6.

\bibitem{Sheikh2020}
F.~Sheikh, Y.~Gao, and T.~Kaiser, ``{A Study of Diffuse Scattering in Massive
  MIMO Channels at Terahertz Frequencies},'' \emph{IEEE Transactions on
  Antennas and Propagation}, vol.~68, no.~2, pp. 997--1008, feb 2020.

\bibitem{Holographic}
A.~Pizzo, T.~L. Marzetta, and L.~Sanguinetti, ``Spatially-stationary model for
  holographic mimo small-scale fading,'' \emph{to appear on JSAC Special Issue
  on Multiple Antenna Technologies for Beyond 5G}, 2020.

\bibitem{Xu2018}
Q.~Xu, C.~Jiang, Y.~Han, B.~Wang, and K.~J. Liu, ``{Waveforming: An Overview
  With Beamforming},'' \emph{IEEE Communications Surveys and Tutorials},
  vol.~20, no.~1, pp. 132--149, 2018.

\bibitem{7982605}
K.~{Lv}, J.~{Hu}, Q.~{Yu}, and K.~{Yang}, ``Throughput maximization and
  fairness assurance in data and energy integrated communication networks,''
  \emph{IEEE Internet of Things Journal}, vol.~5, no.~2, pp. 636--644, April
  2018.

\bibitem{10.1145/3300061.3345451}
A.~Varshney, A.~Soleiman, and T.~Voigt, ``{TunnelScatter: Low Power
  Communication for Sensor Tags Using Tunnel Diodes},'' in \emph{The 25th
  Annual International Conference on Mobile Computing and Networking}.\hskip
  1em plus 0.5em minus 0.4em\relax Association for Computing Machinery, 2019.

\bibitem{NutVLC}
J.~{Beysens}, Q.~{Wang}, and S.~{Pollin}, ``Improving blockage robustness in
  vlc networks,'' in \emph{In Proceedings of the International Conference on
  Communication Systems Networks (COMSNETS)}, 2019, pp. 164--171.

\end{thebibliography}

\end{document}